\documentclass[aps,preprint]{revtex4}
\usepackage{epsfig}
\usepackage{amsmath}
\usepackage{epsfig}

\begin{document}

\title
{\bf Reciprocity and unitarity in  scattering from
a non-Hermitian complex PT-symmetric potential}    
\author{Zafar Ahmed}
\email{zahmed@barc.gov.in}
\affiliation{Nuclear Physics Division, Bhabha Atomic Research Centre,
Mumbai 400 085, India}

\date{\today}

\begin{abstract}
In quantum scattering, Hermiticity is
necessary for both reciprocity and unitarity.  Reciprocity means that both reflectivity ($R$) and transmitivity ($T$) are insensitive to the  direction of incidence of a wave (particle) at a scatterer from left/right.
Unitarity means that $R+T=1$. In scattering from non-Hermitian PT-symmetric structures the (left/right) handedness (non-reciprocity) of reflectivity is known to be essential
and unitarity remains elusive so far. Here we present a surprising occurrence of both reciprocity and unitarity
in some parametric regimes of scattering from a complex PT-symmetric potential.
In special cases, we show that this potential can even become invisible $(R=0, T=1)$ remarkably this time from both left and right sides.
We also find that this potential in a parametric regime enjoys a pseudo-unitarity of the type: $T+\sqrt{R_{left} R_{right}}=1$. 
  
\end{abstract}

\maketitle
In non-relativistic quantum mechanics  Hermiticity is the necessary condition for a Hamiltonian to have: real discrete spectrum, and both  unitarity and  reciprocity in scattering. Reciprocity means that both reflectivity $(R)$ and transmitivity  $(T)$ are insensitive to the  direction of incidence of a wave (particle) at a scatterer from left/right. Unitarity in scattering means that $R+T=1$.
In various branches of physics the complex optical potentials have been in use since a long time to account for the absorption of the incident flux in to unknown channels. Consequently, non-Hermiticity is synonymous to absorption or emission of flux. In this kind of scattering  the unitarity is broken as the probability of reflection ($R$) and transmission ($T$) do not add to 1 and one instead has $R+T+A=1$ where $A$ is the probability of absorption. 

Bender and Boettcher [1,2] conjectured that the eigenspectrum of a non-Hermitian complex potential in a parametric regime was discrete and real. This potential was PT-symmetric [invariant under Parity $(x\rightarrow -x)$ and Time-reversal $(i \rightarrow -i)$].  Also this potential was not amenable to 
exact analytic solutions so it required special methods
to prove the reality of its  spectrum [3]. 
Their conjecture has initiated a debate: `Must a Hamiltonian be Hermitian ?'[2] and it has inspired a large body of investigations leading to the extension of the quantum mechanics in complex domain(see e.g., [1-21,23-31]). About thirteen years later, the present work addresses the same question but this time for reciprocity and unitarity in scattering. Surprisingly, this time again the answer is no.

For the scattering from a complex non-Hermitian potential it has been possible to prove [4] that if  non-Hermitian 
complex potential is spatially asymmetric the reflectivity (R) shows handedness $R_{left} \ne R_{
right}$ whereas transmitivity (T) remains invariant to the direction of the incidence of the particle from left or right. The complex PT-symmetric potentials being spatially anti-symmetric are automatically entitled to this handedness [5-18]. This contrasting feature of scattering from that of reciprocity in Hermitian case perhaps may have discouraged one to look  for unitarity in the scattering from complex PT-symmetric potentials. Nevertheless, various works [4-18] normally display non-unitarity in scattering from complex PT-symmetric potentials.

There has been a very impressive  progress in the investigations of the scattering from a complex PT-symmetric potential. In some PT-symmetric structures  the absence unitarity has been marked with new pseudo-unitarity conditions such as  
$T-1=\pm\sqrt{R_{left}R_{right}}, {R_{left}+R_{right} \over 2}-T=1$(see Eqs. (9) and (17), respectively in Ref. [9]).
The concepts like  spectral singularity [11-14] and invisibility [15,18] have been well developed
both theoretically and experimentally. For the spectral singularity  one looks for positive energies where there are very large (infinite) [11] peaks in both $R(k)$ and $T(k)$. The instance where $R_{right}(k)$ or $R_{left}(k)=0$ and $T(k)=1$ is called unidirectional invisibility [15-18] in both complex-non-Hermitian and complex PT-symmetric potentials. Notice that this invisibility is direction dependent either from left or from right. This is the consequence of the handedness of reflectivity in these potentials.

The ghost of non-Hermiticity has already been busted in PT-symmetric domain, new features such as spectral singularity [11-14] and invisibility [15-17] of such potentials are  being investigated. Novel optical devices and materials  have been engineered to realize wave propagation through a complex PT-symmetric medium [9,15-17,19-21].
In this scenario of scattering from a complex PT-symmetric
potential, we present surprising parametric regimes  in the well known complex PT-symmetric Scarf II potential wherein we observe reciprocity, unitarity, invisibility (from both  left and right). We also find that this potential satisfies one of the newly proposed pseudo-unitarity conditions [9].

The  scattering from Scarf II
potential is well studied wherein the reflection and transmission amplitudes have been well worked out [5,22].
The non-Hermitian complex PT-symmetric  version of the Scarf II potential has been very useful in the investigations of
complex PT-symmetric potentials in various 
ways [5,7,13,14,23-30]. We would like to write the complex Scarf II potential as
\begin{equation}
V(x)=-(B^2+A^2+A)\mbox{sech}^2 x + iB(2A+1) \tanh x~ \mbox{sech}x 
\end{equation}
which is known to have real discrete spectrum [5,22-28].
This potential in another parametric form displays
phase-transition [24] of real discrete eigenvalues
to complex conjugate pairs about a critical value of a parameter when the PT-symmetry breaks down [1,2].
Let $2\mu=1=\hbar^2$ and $k=\sqrt{E}$, where $E$ is the energy. Following [5,22], we can write the transmission amplitude for (1) as [30]
\begin{equation}
t_{A,B}(k)=\frac{\Gamma[-A-ik] \Gamma[1+A-ik] \Gamma[\frac{1}{2}+B-ik] \Gamma[\frac{1}{2}-B-ik]}{\Gamma[-ik] \Gamma[1-ik] \Gamma^2[\frac{1}{2}-ik]},
\end{equation}
\begin{equation}
r_{A,B}(k)=t_{A,B}(k) i\left [\frac{\cos \pi A \sin \pi B}{\cosh \pi k}+  \frac{\sin \pi A \cos \pi B}{\sinh \pi k} \right ].
\end{equation}
The transmitivity $T(k)=|t(k)|^2$ and the reflectivity $R(k)=|r(k)|^2$.
We have re-derived (2,3) to find [14] that for (1)
\begin{eqnarray}
&&t_{left}(k)=t_{A,B}(k), r_{left}(k)=r_{A,B}(k)\\ \nonumber
&&\mbox{and} \quad t_{right}(k)=t_{A,-B}(k), r_{right}(k)=r_{A,-B}(k).
\end{eqnarray}
However, this point can also be verified easily by noticing that the potential (1) satisfies: $V(x,B)=V(-x,-B)$.
Making multiple use of  the property of Gamma functions namely $\Gamma(z) \Gamma(1-z)= \pi ~ \mbox{cosec} \pi z$
we express the transmitivity, $T(k)$ as 
\begin{equation}
T(k)={\sinh^2 \pi k \cosh^2 \pi k \over (\sinh^2\pi k+ \sin^2 \pi A) (\sinh^2 \pi k+ \cos^ 2 \pi B)}, \quad A,B \in R
\end{equation}
It follows that $T(k)$ will be both normal $(<1)$ and anomalous $(>1)$. For the cases $A=B$ or when $A=n+1/2$, or $B=n$, $n \in I$, the transmitivity is normal. 
For the cases  $A=n$ and $A \ne B$ the transmitivity is anomalous at small
energies. Moreover, when the transmitivity (5) is normal, it can be readily checked that the present results on $T(E)$ and $R_{left}(E)$ and $R_{right}(E)$ (2-5) satisfy a pseudo-unitarity of the type:
\begin{equation}
T(E)+\sqrt{R_{left}(E) R_{right}(E)}=1,
\end{equation}
See Ref. [9] for this (6) and other proposals of pseudo-unitarity. 
Ordinarily, for real values of the parameters $A,B$
the Eqs. (2-5) yield to the rule [4,8,9,11,15,16] of left/right handedness (non-reciprocity) of $R(k)$ and sometimes  the  non-unitarity manifests  as a pseudo-unitarity condition (6). For other (in)variances see [14]. 
When $A=-(n+1)-i\alpha$ and $B=i\alpha-(n+1/2)$ with $n\in I^+ +\{0\}, \alpha>0$ in Eqs. (2,3) a recent phenomenon of spectral singularity [11] is observed wherein at $E=\alpha^2$ [14] both $R$ and $T$ become infinite. For other special values of the parameters $A$ and $B$, from Eqs. (2-5) the following  extra-ordinary features arise: \\ \\
{\bf \{1\} Reciprocity and unitarity}\\ 
{\bf Case 1:}\\
When $A=n+1/2, n \in I $ and $B$ is real, from (2) and (3) we get 
\begin{equation}
R(k) ={\cos^2 \pi B \over \sinh^2 \pi k+ \cos^2 \pi B},
\quad 
T(k) ={\sinh^2  \pi k \over \sinh^2 \pi k+ \cos^2 \pi B},
\end{equation}
{\bf Case 2:}\\
When $B \in I$ and $A$ is real,we get  
\begin{equation}
R(k) ={\sin^2 \pi A \over \sinh^2 \pi k+ \sin^2 \pi A}, 
\quad
T(k) ={\sinh^2 \pi k \over \sinh^2 \pi k+ \sin^2 \pi A},
\end{equation} \\
In both the cases from Eq.(4) the acclaimed reciprocity of reflectivity follows: $R_{left}(k)=R_{right}(k)$. The reflectivity is also symmetric under time-reversal:
$R(-k)=R(k)$. The claimed unitarity can be checked readily using Eqs.(7,8). $\diamond$ \\ \\
{\bf \{2\} Invisibility with reciprocity:}\\  
In the above two cases of unitarity  when $A=(n+1/2), B=(m+1/2)$ or $A=n, B=m$ ($ n,m \in I$) check that two cases of invisibility occur wherein $R(k)=0, T(k)=1$ at any energy. This invisibility of the complex PT-symmetric potential (1) is not unidirectional [15-18], this time it is from both sides left and right.  $\diamond$ \\

Earlier such an invisibility has been termed as bi-directional invisibility and first found in $V(x)=-{1 \over (x+ia)^2}$ [31] using the methods super symmetric 
quantum mechanics. More recently bi-directional invisibility has been found in the PT-symmetric Ginocchio's potential [32]. Interestingly, both these instances are for  the potentials of the type $V(x)=V_0 f(x+ia)$ where a real Hermitian potential has been complexified by an imaginary shift of the $x$ co-ordinate. In these types of potentials as argued in [5] $R_{left}, R_{right}$
are of the type $e^{\pm 2ka} R^\prime$  where $R^\prime$
is reflectivity of the Hermitian potential $(V_0 f(x))$.
Consequently, the pseudo-unitarity (6) is satisfied and
the left and right reflectivity zero(s) occur at the same discrete energ(ies).

The reflectivity of the non-Hermitian complex potentials which are spatially symmetric shows [4]  reciprocity
along with non-unitarity $ R+T <(>) 1$ when the imaginary
part of the potential is negative (positive) definite for $x \in (-\infty, \infty)$. For example for $V(x)=(V_1\mp iV_2)~ \mbox{sech}^2 x, V_1, V_2 \in R, V_2>0~ (V_2 <0)$, the reflectivity will show reciprocity along with non-unitarity  $R+T<(>)1$, respectively.
Hence, we  know that reciprocity does not imply unitarity. We speculate that unitarity may be sufficient for reciprocity of reflectivity in scattering from a complex   non-Hermitian potential.

Notwithstanding a rapid research in PT-symmetric structures these days and the familiarity of the complex PT-symmetric Scarf II potential, to the best of our knowledge, the above two paradoxical
or exceptional features $\{1,2\}$  are new and un-noticed, so far. The  observations and the proofs of the non-reciprocity of reflectivity in scattering from complex PT-symmetric potentials are abound, however, one ought to look for reciprocity under some special parametric  condition hereafter. 
The question whether there are various parametric regimes  in other PT-symmetric structures yielding to  reciprocity (of reflectivity), unitarity, spectral singularity,  invisibility and the pseudo-unitarity of the type (6) is open for investigations. With regard to this, studying exactly solvable complex potentials becomes even more important. The present exposition provides a paradigm shift in the thinking in two ways. Firstly, in scattering,
Hermiticity and time-reversal symmetry of an interaction are not necessary for unitarity and reciprocity (of reflectivity), respectively. Secondly, the complex PT-symmetric structures are very versatile having multiple parametric regimes displaying various properties.

\section*{References}

\begin{enumerate}
\item C. M. Bender and S. Boettcher, Phys. Rev. Lett. {\bf 80} (1998) 5243.
\item C. M. Bender, D.C. Brody, H.F. Jones
, Am. J. Phys {\bf 71} (2003) 1095.
\item P. Dorey, C. Dunning and R. Tateo, J. Phys. A: Math. \& Gen. {\bf 34} (2001) 5679.
\item Z. Ahmed,  Phys. Rev. A {\bf 64} (2001) 042716.
\item G. Levai, F. Cannata and A. Ventura, J. Phys. A: Math. Gen. {\bf 34} (2001) 839.
\item R. N. Deb, A. Khare , B.D. Roy, Phys. Lett A {\bf 307} (2003) 215.  
\item Z. Ahmed, Phys. Lett. A {\bf 324} (2004) 152.
\item F. Cannata, J.-P. Dedonder and A. Ventura, Ann. Phys.(N.Y.) {\bf 322} (2007) 397.
\item Li Ge, Y.D. Chong and A.D. Stone ,  
arXiv 1112-5167v1 [physics-optics] Dec 21. 2011
(See also refs. there in)
\item H. F. Jones,  J. Phys. A: Math. Theor. {\bf 45} (2011) 135306.
\item A. Mostafazadeh, Phys. Rev. Lett. {\bf 102} (2009) 220402. 
\item S. Longhi, Phys. Rev. A {\bf 81} (2010) 022102.
\item Z. Ahmed, J. Phys. A: Math. Theor. {\bf 42} (2009) 472005.
\item Z. Ahmed 2012, J. Phys. A: Math. Theor. {\bf 45} (2009) 032004.
\item M. Kulishov, J. M. Laniel, N. Belanger, J. Azana  and D. V. Plant, Opt. Express {\bf 13} (2005) 3068.
\item Z. Lin, H. Ramezani,  T. Eichelkraut, T. Kottos,  H. Cao and D.N. Christodoulides,  Phys. Rev. Lett. {\bf 106} (2011) 213901.
\item S. Longhi, J. Phys. A: Math. Theor. 44 (2011) 485302.
\item A. Mostafazadeh,  arXiv: 1206.0116v1 [mat-ph] 1 June, 2012.
\item Z. H. Musslimani, K.G. Makris, R. El-Ganainy,  and D.N. Christodoulides, Phys. Rev. Lett. {\bf 100} (2008) 030402. (See also references there in)
\item A. Guo, G. J. Salamo,D. Duchesne, R. Morandotti, M. Volatier-Ravat, V. Aimez, G. A. Siviloglou and D. N. Christodoulides, Phys. Rev. Lett. {\bf 103} (2009) 093902.
\item C. E. R{\"u}ter, G.E. Makris,R. El-Ganainy, D.N.  
Christodoulides, M. Segev and D. Kip,  Nature Physics {\bf 6} (2010) 192. 
\item A. Khare, and U.P. Sukhatme, J. Phys. A: Math. Gen. {\bf 21} (1988) L501. 
\item B. Bagchi, C. Quesne,  Phys. Lett A: {\bf 273} (2000) 285.
\item Z. Ahmed, Phys. Lett.  A {\bf 282} (2001)343-348; {\bf 287} 295. 
\item  G. Levai and M. Znojil, J. Phys. A: Math. Gen. {\bf 35} (2002) 8793.
\item  G. Levai, F. Cannata and A. Ventura, J. Phys. A: Math. Gen. {\bf 35} (2002) 5041.
\item  G. Levai, F. Cannata and A. Ventura, Phys. Lett. A {\bf 300} (2002) 271.
\item F. Correa and M.S. Plyushchay, Ann. Phys. (N.Y) {\bf 327} (2012) 1761.
\item A. Sinha, Euro. Phys. Lett. {\bf 98} (2012) 60005.
\item These Eqs. are the same as given in (Eq. (6) in Ref. [14]), excepting an error for $r_{A,B}$ in [14]. There [14] $\sin \pi B$ and $\cos \pi B$ should have appeared correctly as $\sinh \pi B$ and $\cosh \pi B$, respectively.
\item A. A. Andrianov, M. V. Ioffe, F. Cannata, J.-P. Dedoner, Int. J. Mod. Phys. A {\bf 14} (2005) 3068.
\item A. Ghatak, A.N. Joseph, B.P. Mandal, Z. Ahmed,
J. Phys. A: Math. Theor. {\bf 45} (2012) 465305.
\end{enumerate}
\end{document}